\documentclass[12pt]{article}
\usepackage{lipsum}
\usepackage[textwidth=17.8cm]{geometry}
\usepackage{times}
\usepackage{color}
\usepackage{amsmath}
\usepackage{amssymb}
\usepackage{graphicx}
\usepackage{bm}
\usepackage[abs]{overpic}
\usepackage{wasysym}
\usepackage{scicite}
\usepackage[affil-it]{authblk}
\def\be{\begin{equation}}
\def\ee{\end{equation}}
\newcommand{\7}{$^7$Li}
\newcommand{\6}{$^6$Li}
\newcommand{\ket}[1]{\left| #1 \right>} 
\newcommand{\bea}{\begin{eqnarray}}
\newcommand{\eea}{\end{eqnarray}}
\graphicspath{{../}}

\usepackage{pdfpages}
\begin{document}


# A Mixture of Bose and Fermi Superfluids

I. Ferrier-Barbut, M. Delehaye, S. Laurent, A. T. Grier†, M. Pierce, B. S. Rem††,
F. Chevy, C. Salomon

Laboratoire Kastler-Brossel, École Normale Supérieure,
Collège de France, CNRS and UPMC,
24 rue Lhomond, 75005 Paris, France

Superconductivity and superfluidity of fermionic and bosonic systems are remarkable many-body quantum phenomena. In liquid helium and dilute gases, Bose and Fermi superfluidity has been observed separately, but producing a mixture in which both the fermionic and the bosonic components are superfluid is challenging. Here we report on the observation of such a mixture with dilute gases of two Lithium isotopes, $^6$Li and $^7$Li. We probe the collective dynamics of this system by exciting center-of-mass oscillations that exhibit extremely low damping below a certain critical velocity. Using high precision spectroscopy of these modes we observe coherent energy exchange and measure the coupling between the two superfluids. Our observations can be captured theoretically using a sum-rule approach that we interpret in terms of two coupled oscillators.

In recent years, ultracold atoms have emerged as a unique tool to engineer and study quantum many-body systems. Examples include weakly interacting Bose-Einstein condensates [1,2], two-dimensional gases [3], and the superfluid-Mott insulator transition [4] in the case of bosonic atoms, and the crossover between Bose-Einstein condensation (BEC) and fermionic superfluidity described by Bardeen, Cooper and Schrieffer's theory (BCS) for fermionic atoms [5]. Mixtures of Bose-Einstein condensates were produced shortly after the observation of BEC

[2] and a BEC mixed with a single spin state Fermi sea was originally observed in [6,7]. However, realizing a mixture where both fermionic and bosonic species are superfluid has been experimentally challenging. This has also been a long-sought goal in liquid helium where superfluidity was achieved separately in both bosonic [4]He and fermionic [3]He. The double superfluid should undergo a transition between s-wave and p-wave Cooper pairs as the [3]He dilution is varied [8]. However, because of strong interactions between the two isotopes, [3]He-[4]He mixtures contain only a small fraction of [3]He (typically 6%) which, so far, has prevented reaching simultaneous superfluidity for the two species [8, 9].

Here we report on the production of a Bose-Fermi mixture of quantum gases where both species are superfluid. Our system is an ultracold gas of fermionic [6]Li in two spin-states mixed with [7]Li bosons and confined in an optical dipole trap. Using radio-frequency pulses [6]Li atoms are prepared in their two lowest hyperfine states $|1_f\rangle$ and $|2_f\rangle$ whereas [7]Li is spin polarized in the second to lowest state $|2_b\rangle$ [10]. For this combination of states, in the vicinity of the [6]Li Feshbach resonance at a magnetic field of 832 Gauss [11], the scattering length of the bosonic isotope $a_b = 70\ a_0$ ($a_0$ is the Bohr radius) is positive, preventing collapse of the BEC. The boson-fermion interaction is characterized by a scattering length $a_{bf} = 40.8\ a_0$ that does not depend on magnetic field in the parameter range studied here. At resonance the Fermi gas exhibits a unitary limited collision rate and lowering the optical dipole trap depth leads to extremely efficient evaporation. Thanks to a large excess of [6]Li atoms with respect to [7]Li, the Bose gas is sympathetically driven to quantum degeneracy.

The two clouds reach the superfluid regime after a 4-second evaporation ramp [10]. As the [7]Li Bose gas is weakly interacting, the onset of BEC is detected by the growth of a narrow

peak in the density profile of the cloud. From previous studies on atomic Bose-Einstein condensates, we conclude that the $^7$Li BEC is in a superfluid phase. Superfluidity in a unitary Fermi gas is notoriously more difficult to detect because of the absence of any qualitative modification of the density profile at the phase transition. To demonstrate the superfluidity of the fermionic component of the cloud, we slightly imbalance the two spin populations. In an imbalanced gas, the cloud is organized in concentric layers, with a fully paired superfluid region at its center where Cooper pairing maintains equal spin populations. This $^6$Li superfluid core can be detected by the presence of a plateau in the doubly integrated density difference [12]. Examples of density profiles of the bosonic and fermionic superfluids are shown in Fig. 1 where both the Bose-Einstein condensate (blue circles) and the plateau (black diamonds in the inset) are clearly visible. Our coldest samples contain $N_b = 4\times10^4$ $^7$Li atoms and $N_f = 3.5\times10^5$ $^6$Li atoms. The absence of a thermal fraction in the bosonic cloud indicates a temperature below $0.5\,T_{c,b}$, where $k_B T_{c,b} = 0.94\hbar\overline{\omega}_b N_b^{1/3}$ is the critical temperature of the $^7$Li bosons, and $\overline{\omega}_b$ ($\overline{\omega}_f$) is the geometric mean trapping frequency for $^7$Li ($^6$Li). Combined with the observation of the $^6$Li plateau, this implies that the Fermi cloud is also superfluid with a temperature below $0.8T_{c,f}$. Here $T_{c,f}$ is the critical temperature for superfluidity of a spin-balanced, harmonically trapped Fermi gas at unitary, $T_{c,f} = 0.19 T_F$ [13] and $k_B T_F = \hbar\overline{\omega}_f(3\,N_f)^{1/3}$ is the Fermi temperature. The superfluid mixture is very stable with a lifetime exceeding 7s for our coldest samples.

As seen in Fig. 1, the Bose-Fermi interaction is too weak to alter significantly the density profiles of the two species [14]. To probe the interaction between the two superfluids we study the dynamics of the mass centers of the two isotopes (dipole modes), a scheme used in the past for the study of mixtures of Bose-Einstein condensates [15, 16], mixtures of Bose-Einstein

condensates and spin-polarized Fermi seas [17], spin diffusion in Fermi gases [18] or integrability in 1D systems [19]. In a purely harmonic trap and in the absence of interspecies interactions, the dipole mode of each species is undamped and can therefore be measured over long time spans to achieve a high frequency resolution and detect small perturbations of the system. We excite the dipole modes by shifting the initial position of the $^6$Li and $^7$Li clouds by a displacement $d$ along the weak direction $z$ of the trap [10]. We then release them and let them evolve during a variable time $t$ after which we measure their positions. Monitoring the cloud oscillations during up to 4 s we determine their frequencies with high precision ($\frac{\Delta\omega}{\omega} \lesssim 2\times10^{-3}$). In the absence of the other species, the oscillation frequencies of $^6$Li and $^7$Li are respectively $\omega_\text{f} = 2\pi\times16.80(2)$ Hz and $\omega_\text{b} = 2\pi\times15.27(1)$ Hz. In the axial direction, the confinement is mostly magnetic and at high magnetic field both species are in the Paschen-Back regime where the electronic and nuclear spin degrees of freedom are decoupled. In this regime the magnetic confinement mostly results from the electronic spin and is therefore almost identical for the two isotopes. The ratio $\omega_\text{f}/\omega_\text{b}$ is then very close to the expected value $\sqrt{7/6} \simeq 1.08$ based on the ratio of the atomic masses [20].

Contrary to the large damping observed in the Bose-Bose mixtures [15], we observe long-lived oscillations of the Bose-Fermi superfluid mixture at frequencies ($\widetilde{\omega}_\text{b}$, $\widetilde{\omega}_\text{f}$). These oscillations extend over more than 4 seconds with undetectable damping (Fig. 2 and Fig. S2). This very weak dissipation is only observed when the initial displacement $d$ is below 100 $\mu$m, corresponding to a maximum relative velocity $v_{max} = (\widetilde{\omega}_\text{b} + \widetilde{\omega}_\text{f})d$ below 18 mm/s $\simeq 0.4\ v_\text{F}$ where $v_\text{F} = \sqrt{2k_\text{B}T_\text{F}/m_\text{f}}$. In this situation the BEC explores only the central part of the much broader Fermi cloud. When $v_{max} > v_\text{c} = 0.42^{+0.05}_{-0.11}v_\text{F} = 20^{+2}_{-5}$ mm/s we observe a sharp onset

of damping and heating of the BEC compatible with the Landau criterion for breakdown of superfuidity (Fig. 2 (c) and [10]). For comparison, the sound velocity of an elongated Fermi gas at its center is $v_s' = \xi^{1/4} v_F/\sqrt{5} = 17$ mm/s [21] where $\xi = 0.38$ is the Bertsch parameter [13, 5]. The measured critical velocity $v_c$ is very close to $v_s'$ and is clearly above the BEC sound velocity of $\simeq 5$ mm/s at its center.

Two striking phenomena are furthermore observed. First, whereas the frequency $\widetilde{\omega}_f$ of $^6$Li oscillations is almost unchanged from the value in the absence of $^7$Li, that of $^7$Li is downshifted to $\widetilde{\omega}_b = 2\pi \times 15.00(2)$ Hz. Second, the amplitude of oscillations of the bosonic species displays a beat at a frequency $\simeq (\widetilde{\omega}_f - \widetilde{\omega}_b)/(2\pi)$ revealing coherent energy transfer between the two clouds (Fig. 2 (b)). To interpret the frequency shift of the $^7$Li atoms, we note that $N_b \ll N_f$, which allows us to treat the BEC as a mesoscopic impurity immersed in a Fermi superfluid. Similarly to the Fermi polaron case [22], the effective potential seen by the bosons is the sum of the trapping potential $V(r)$ and the mean-field interaction $g_{bf} n_f(r)$, where $n_f$ is the total fermion density, $g_{bf} = 2\pi \hbar^2 a_{bf}/m_{bf}$, and $m_{bf} = \frac{m_b m_f}{m_b + m_f}$ is the $^6$Li/$^7$Li reduced mass. Neglecting at first the back action of the bosons on the fermions, we can assume that $n_f$ is given by the Local-Density-Approximation result $n_f(r) = n_f^{(0)}(\mu_f^0 - V(r))$, where $n_f^{(0)}(\mu)$ is the stationary equation of state of the Fermi gas (EoS). Because the Bose-Einstein condensate is much smaller than the Fermi cloud (Fig. 2 (a)), $V(r)$ is smaller than $\mu_f^0$ over the BEC volume. We can thus expand $n_f^{(0)}$ and we get

$$V_{\text{eff}}(r) = g_{bf} n_f(0) + V(r)\left[1 - g_{bf}\left(\frac{dn_f^{(0)}}{d\mu_f}\right)_{r=0}\right]. \qquad (1)$$

We observe that the effective potential is still harmonic and the rescaled frequency is given by

$$\tilde{\omega}_b \simeq \omega_b \left(1 - \frac{1}{2} g_{bf} \left(\frac{dn_f^{(0)}}{d\mu_f}\right)_{r=0}\right). \tag{2}$$

For a unitary Fermi gas, the chemical potential is related to the density by $\mu_f = \xi \hbar^2 (3\pi^2 n_f)^{2/3}/2m_f$. In the weakly coupled limit we get $\frac{\delta \omega_b}{\omega_b} = \frac{\omega_b - \tilde{\omega}_b}{\omega_b} = \frac{13\, k_F a_{bf}}{7\pi \xi^{5/4}}$, where $\hbar k_F = \sqrt{2\hbar m_f \bar{\omega}_f (3N_f)^{1/3}}$ is the Fermi momentum of a non-interacting harmonically trapped Fermi gas. Using our experimental parameters $k_F = 4.6 \times 10^6$ m$^{-1}$, we predict a value $\tilde{\omega}_b \simeq 2\pi \times 14.97$ Hz, in very good agreement with the observed value 15.00(2)Hz.

To understand the amplitude modulation, we now take into account the back-action on the fermions. A fully quantum formalism using a sum-rule approach [23, 24, 25] leads to a coupled oscillator model where the positions of the two clouds obey the following equations [10]

$$M_f \ddot{z}_f = -K_f z_f - K_{bf}(z_f - z_b) \tag{4}$$

$$M_b \ddot{z}_b = -K_b z_b - K_{bf}(z_b - z_f), \tag{5}$$

where $M_b = N_b m_b$ ($M_f = N_f m_f$) is the total mass of the $^7$Li ($^6$Li) cloud, $K_b = M_b \omega_b^2$ ($K_f = M_f \omega_f^2$) is the spring constant of the axial magnetic confinement, and $K_{bf}$ is a phenomenological (weak) coupling constant describing the mean-field interaction between the two isotopes. To recover the correct frequency shift (Eq. 2), we take $K_{bf} = 2K_b \frac{\delta \omega_b}{\omega_b}$.

Solving these equations with the initial condition $z_f(0) = z_b(0) = d$ and defining $\rho = N_b/N_f$ and $\varepsilon = \frac{2m_b}{m_b - m_f} \left(\frac{\tilde{\omega}_b - \omega_b}{\omega_b}\right)$, in the limit $\rho, \varepsilon \ll 1$ we get:

$$z_f = d[(1 - \varepsilon\rho)\cos(\tilde{\omega}_f t) + \varepsilon\rho \cos(\tilde{\omega}_b t)] \tag{7}$$

$$z_b = d[-\varepsilon \cos(\tilde{\omega}_f t) + (1 + \varepsilon)\cos(\tilde{\omega}_b t)]. \tag{8}$$

The predictions of Eqs. 7 and 8 agree well with experiment (Fig. 2 (b)). Interestingly, the

peak-to-peak modulation of the amplitude of $^7$Li is much larger than the relative frequency shift, a consequence of the almost exact tuning of the two oscillators (up to a factor $\sqrt{6/7}$). Thus the mass pre-factor in the expression for $\varepsilon$ is large (=14) and leads to $\varepsilon \simeq 0.25$ at unitarity. This results in efficient energy transfer between the two modes despite their weak coupling, as observed.

We now extend our study of the Bose-Fermi superfluid mixture to the BEC-BCS crossover by tuning the magnetic field away from the resonance value $B_f = 832$ G. We explore a region from 860 G down to 780 G were $1/k_F a_f$ spans the interval $[-0.4, +0.8]$. In this whole domain, except in a narrow region between 845 G and 850 G where the boson-boson scattering length is negative, the mixture is stable and the damping extremely small.

The frequency shift of the BEC (Eq. 2) now probes the derivative of the equation of state $n_f(\mu_f)$ in the BEC-BCS crossover. In the zero-temperature limit and under the Local Density Approximation, Eq. (2) obeys the universal scaling $\frac{\delta \omega_b}{\omega_b} = k_F a_{bf} f\left(\frac{1}{k_F a_f}\right)$.

In Fig. 3 we compare our measurements to the prediction for the function $f$ obtained from the zero-temperature EoS measured in [26]. On the BCS side, ($1/k_F a_f < 0$) the frequency shift is reduced and tends to that of a non-interacting Fermi gas. Far on the BEC side ($1/k_F a_f \gg 1$) we can compute the frequency shift using the EoS of a weakly interacting gas of dimers. Within the mean-field approximation we have $\frac{dn_f}{d\mu_f} = \frac{2\, m_f}{\pi \hbar^2 a_{dd}}$ where $a_{dd} = 0.6 a_f$ is the dimer-dimer scattering length. This expression explains the increase of the frequency shift when $a_f$ is reduced, i.e moving towards the BEC side (See [10] for the effect of Lee-Huang-Yang quantum correction).

The excellent agreement between experiment and our model confirms that precision measurements of collective modes are a sensitive dynamical probe of equilibrium properties of many-body quantum systems [27]. Our approach can be extended to the study of higher-order excitations. In particular, while there are two first sound modes, one for each atomic species, we expect only one second sound for the superfluid mixture [28] if cross-thermalization is fast enough. In addition, the origin of the critical velocity for the relative motion of Bose and Fermi superfluids is an intriguing question that can be further explored in our system. Finally, a richer phase diagram may be revealed when $N_b/N_f$ is increased [29] or when the superfluid mixture is loaded in an optical lattice [30].

† Present address: Van Swinderen Institute, FMNS, University of Groningen, Zernikelaan 25, 9747 AA Groningen, Netherlands

†† Present address: Institut für Laserphysik, Universität Hamburg, Luruper Chaussee 149, Building 69, D-22761 Hamburg, Germany

## Figures

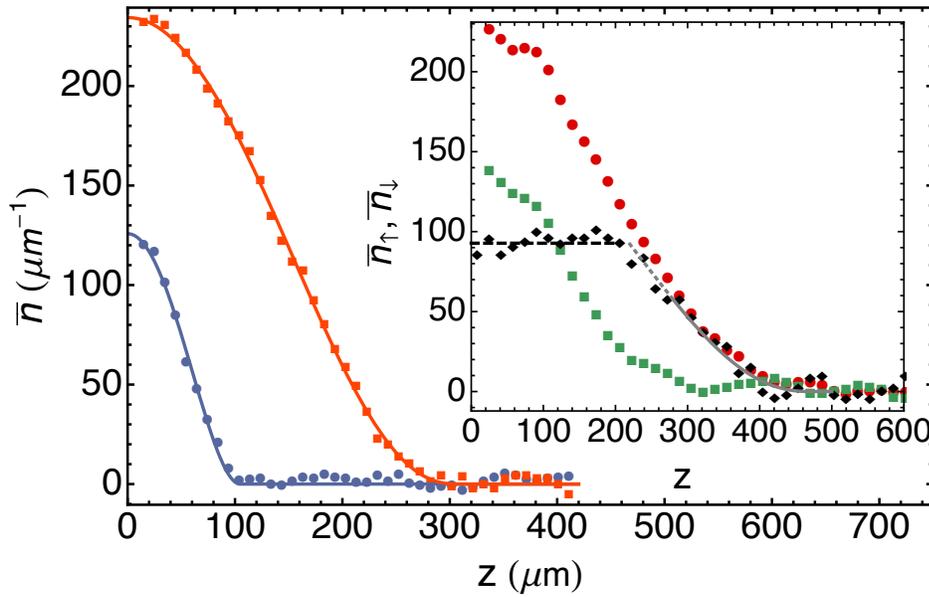

- **Figure 1**: **Density profiles in the double superfluid regime.** $N_\mathrm{b} = 4\times10^4$ $^7$Li atoms and $N_\mathrm{f} = 3.5\times10^5$ $^6$Li atoms are confined in a trap at a temperature below 130 nK. The density profiles $\bar{n}_\mathrm{b}$ (blue circles) and $\bar{n}_{\mathrm{f},\uparrow}$ (red squares) are doubly integrated over the two transverse directions. The blue (red) solid line is a fit to the of $^7$Li ($^6$Li) distribution by a mean-field (unitary Fermi gas) equation of state in the Thomas-Fermi approximation. Inset: Spin-imbalanced Fermi gas ($N_{\mathrm{f},\uparrow} = 2\times10^5$, $N_{\mathrm{f},\downarrow} = 8\times10^4$) in thermal equilibrium with a BEC. Red circles: $\bar{n}_{\mathrm{f},\uparrow}$ ; Green

squares: $\bar{n}_{f,\downarrow}$; Black diamonds : difference $\bar{n}_{f,\uparrow} - \bar{n}_{f,\downarrow}$. The plateau (black dashed line) indicates superfluid pairing [12]. Grey solid line : Thomas-Fermi profile of a non-interacting Fermi gas for the majority component outer shell.

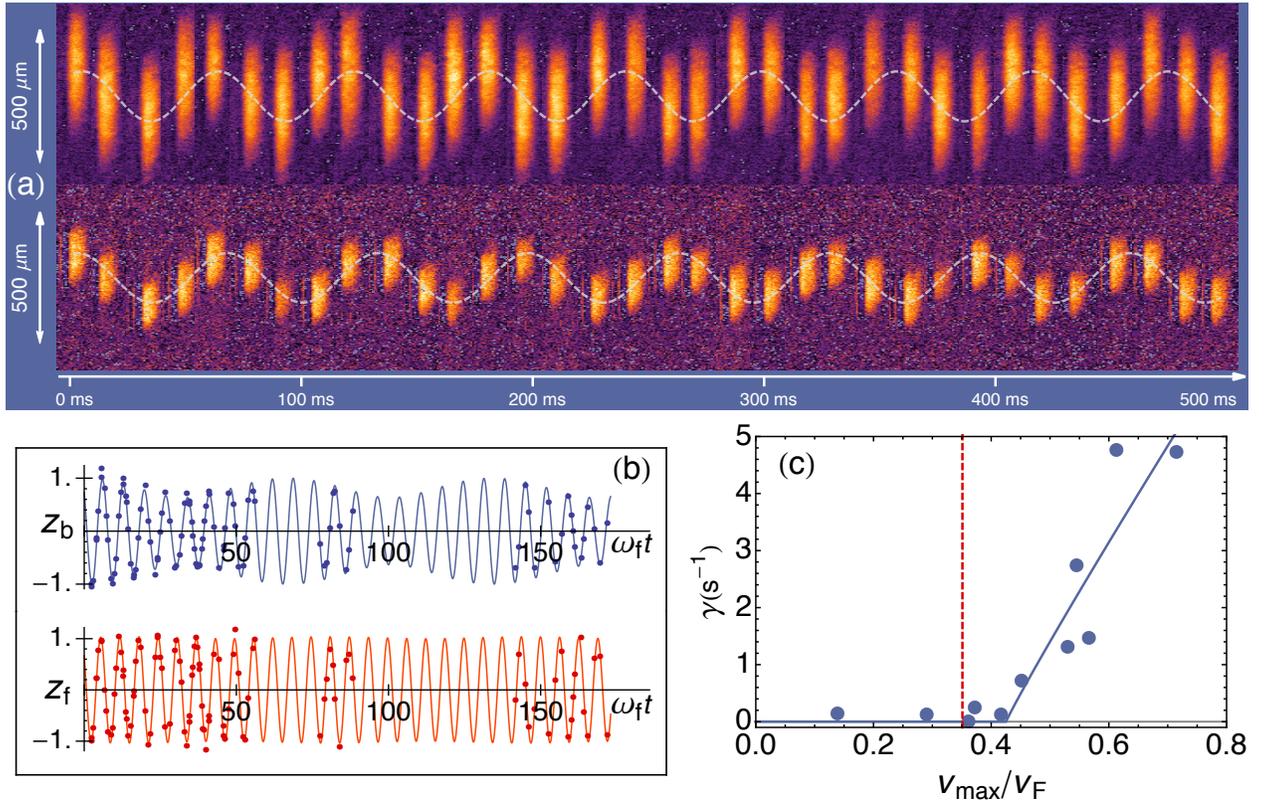

- **Figure 2**: **(a) Center of mass oscillations**. The oscillations are shown over the first 500 ms at a magnetic field of 835 G for a Fermi superfluid (top) and a Bose superfluid (bottom). The oscillation period of $^6$Li ($^7$Li) is 59.7(1) ms (66.6(1) ms), leading to a dephasing of $\pi$ near 300 ms. These oscillations persist for more than 4s with no visible damping. The maximum relative velocity between the two clouds is 1.8 cm/s.

  **(b) Coupled oscillations.** Symbols: Center-of-mass oscillation of $^7$Li (top) and $^6$Li (bottom) displaying coherent energy exchange between both superfluids. Solid lines: Theory for an initial displacement $d$ of 100 $\mu$m at a magnetic field of 835 G, see text.

**(c) Critical Damping.** Symbols: Damping rate (blue circles) of the amplitude of the center-of-mass oscillations of the $^7$Li BEC as a function of the maximal relative velocity between the two superfluids normalized to the Fermi velocity of the $^6$Li gas. Data taken at 832 G. From this data and using a fit function given in [10] (solid line) we extract $v_c = 0.42^{+0.05}_{-0.11} v_F$. The red dashed line shows the speed of sound of an elongated unitary Fermi superfluid $v'_s = \xi^{1/4} v_F / \sqrt{5} = 0.35 v_F$ [20].

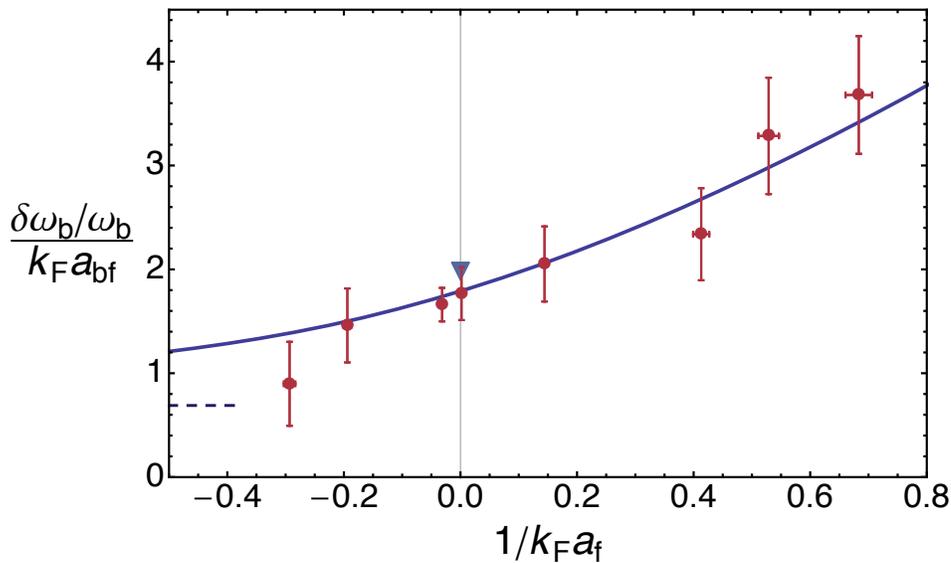

- **Figure 3: Dipole mode frequency shift in the BEC-BCS crossover.** Red circles: Experiment. Blue line: zero-temperature prediction from the Equation of State of ref [26]; Dashed line: ideal Fermi gas. Blue triangle, prediction from ref [13].


\title{\vspace{-1.cm}\Large{\bf Supplementary material to\\ A Mixture of Bose and Fermi Superfluids}}
\author{\normalsize I.~Ferrier-Barbut}
\author{\normalsize M.~Delehaye}
\author{\normalsize S.~Laurent}
\author{\normalsize A.~T.~Grier}
\author{\normalsize M.~Pierce}
\author{\normalsize B. S.~Rem}
\author{\normalsize F.~Chevy}
\author{\normalsize C.~Salomon}
\affil{\normalsize \vspace{-.35cm}Laboratoire Kastler-Brossel, \'Ecole Normale Sup\'erieure,\\ Coll\`ege de France, CNRS and UPMC, 24 rue Lhomond, 75005 Paris, France}
\date{\vspace{-5ex}}
\maketitle
%

{\bf{\large Feshbach Resonances}}\\

\begin{figure}[h!]
\centerline{
\includegraphics[width=.6\columnwidth]{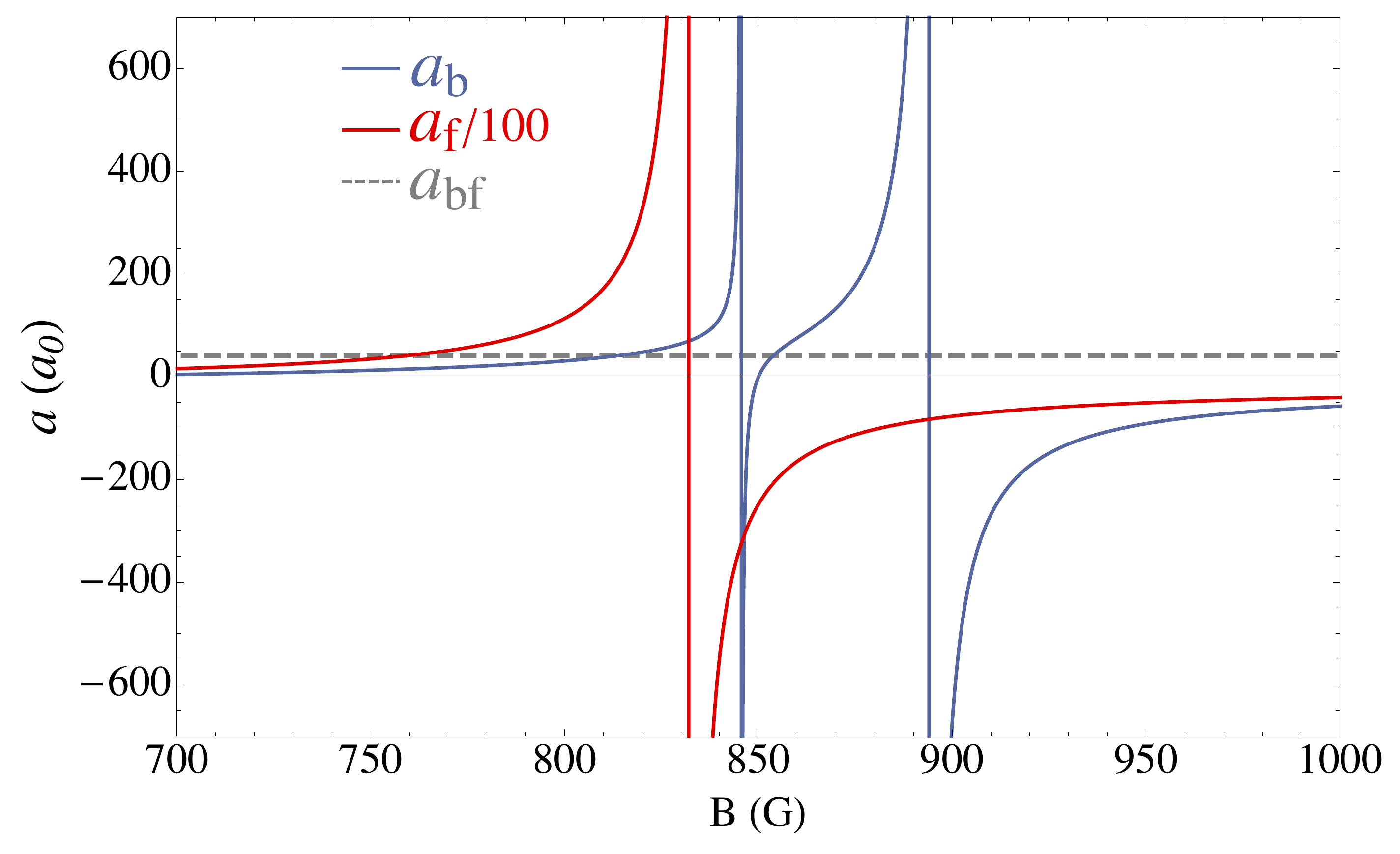}}
\caption{Magnetic field dependence of the different scattering lengths $a_{\rm b}$ (blue), $a_{\rm f}$ (red), and $a_{\rm bf}$ (dashed gray). $a_{\rm bf}= 40.8\,a_0$ is independent of $B$. Note the 1/100 vertical scale for \6.   }
\label{fig:FR}

\end{figure}

The Bose-Fermi mixture is composed of a  \7 cloud prepared in the $\ket{2_{\rm b}}$ state, which connects to the $\ket{F=1,m_{\rm f}=0}$ state at low field, together with a \6 gas in the two lowest energy states $\ket{1_{\rm f}}$ and $\ket{2_{\rm f}}$ connecting to $\ket{F=1/2,m_{\rm f}=1/2}$ and $\ket{F=1/2,m_{\rm f}=-1/2}$ respectively. \\
In Fig.~\ref{fig:FR} we present the relevant s-wave scattering lengths characterizing the \7-\7, \6-\6 and \6-\7 interactions in the 700\,G-1000\,G magnetic field region of interest.
 \7, $\ket{2_{\rm b}}$ exhibits two Feshbach resonances located  at 845.5\,G and 894\,G.
For fermionic \6, the two spin-states $\ket{1_{\rm f}}$, $\ket{2_{\rm f}}$ exhibit one very broad s-wave resonance at 832.18\,G. Note the 1/100 vertical scale for \6 in Fig.~\ref{fig:FR}. The scattering lengths are taken from ({\it 11,31}) in units of Bohr radius $a_0$ as a function of magnetic field $B$ in gauss:
\footnotesize
\bea
a_{\rm f}(B)&=&-1582\left(1-\frac{-262.3}{B-832.18}\right)\\
a_{\rm b}(B)&=&-18.24\left(1-\frac{-237.8}{B-893.95}\right)\left(1-\frac{4.518}{B-845.54}\right)
\eea
\normalsize
For the inter-isotope interaction,  coupled-channel calculations by S. Kokkelmans provide  a scattering length $a_{\rm bf}=40.8\,a_0$ independent of the magnetic field in this region.

\newpage
{\bf{\large Experimental set-up, mixture preparation}}\\

The apparatus and early stages of our experiment have been described in \cite{nascimbene2009pol}. Initially \7 (resp. \6) atoms are cooled to $40\,\mu$K in a Ioffe-Pritchard trap in the $\ket{F=2,m_{\rm f}=2}$ (resp. $\ket{F=3/2,m_{\rm f}=3/2}$) states at a bias field of 12.9\,G. The trapping potential for the mixture is a hybrid trap composed of an optical dipole trap (wavelength 1.07\,$\mu m$) with waist 27$\mu$m superimposed with a magnetic curvature in which the bias magnetic field remains freely adjustable.
About $3\times 10^5$ \7 and $2\times 10^6$ \6 atoms are transferred in a 300\,$\mu$K deep optical dipole trap. They are then transferred to their absolute ground state $\ket{F=1,m_{\rm f}=1}$ and $\ket{F=1/2,m_{\rm f}=1/2}$ by a rapid adiabatic passage (RAP) using two 50\,ms radio-frequency (RF) pulses and a sweep of the magnetic bias down to 4.3\,G. These states connect at high magnetic field respectively to $\ket{1_{\rm b}}$ and $\ket{1_{\rm f}}$. We revert the magnetic curvature in order to provide an axial confining potential, the ground states being high-field-seeking states. The bias field is ramped in 100\,ms to 656\,G where we transfer \7 to the state $\ket{2_{\rm b}}$ by a RAP done by an RF pulse with a  frequency sweep from 170.9\,MHz to 170.7\,MHz in 5\,ms.
The field is ramped in 100\,ms  to 835\,G where a mixture of \6 in its two lowest energy states  $\ket{1_{\rm f}}$ and  $\ket{2_{\rm f}}$ is prepared with an RF sweep between 76.35\,MHz and 76.25\,MHz. The duration of this sweep varies the Landau-Zener efficiency of the transfer offering control of the spin polarization of the \6 mixture. Initial conditions for evaporation at this field are $1.5\times10^5$ \7 and $1.5\times10^6$ \6 at 30\,$\mu$K in a 300\,$\mu$K deep trap.
The evaporation of the mixture is done near unitarity for the fermions providing high collision rate. In 3\,s the laser power is reduced by a factor 100 and \7 is sympathetically cooled by \6 with high efficiency; the phase-space density increases to BEC by a factor $\sim2\times10^4$ for a factor of ten loss in \7 atoms. To confirm this sympathetic cooling scheme we have also performed the evaporation at 850~G where the \7 scattering length vanishes, demonstrating that \7 can be cooled down solely by thermalisation with \6. At the end of evaporation, we typically wait 700\,ms at constant dipole trap power to ensure thermal equilibrium between both species.\\

The trapping potential is cylindrically symmetric, with axial (transverse) frequency $\omega_z$ ($\omega_\rho$). The BEC phase transition is observed at a temperature of 700~nK.
Our studies are performed in a shallow trap with frequencies:
\begin{itemize}
\item$\omega_{\rho,\rm b}=2\pi\times550(20)$~Hz, $\omega_{\rho,\rm f}=2\pi\times595(20)$~Hz
\item$\omega_{z,\rm b}=2\pi\times15.27$~Hz, $\omega_{z,\rm f}=2\pi\times16.8$~Hz.
\end{itemize}
These frequencies are measured by single species center-of-mass oscillations at a field of 832G.

\begin{figure}[htbp]
\centerline{
\begin{overpic}[width=.6\columnwidth]{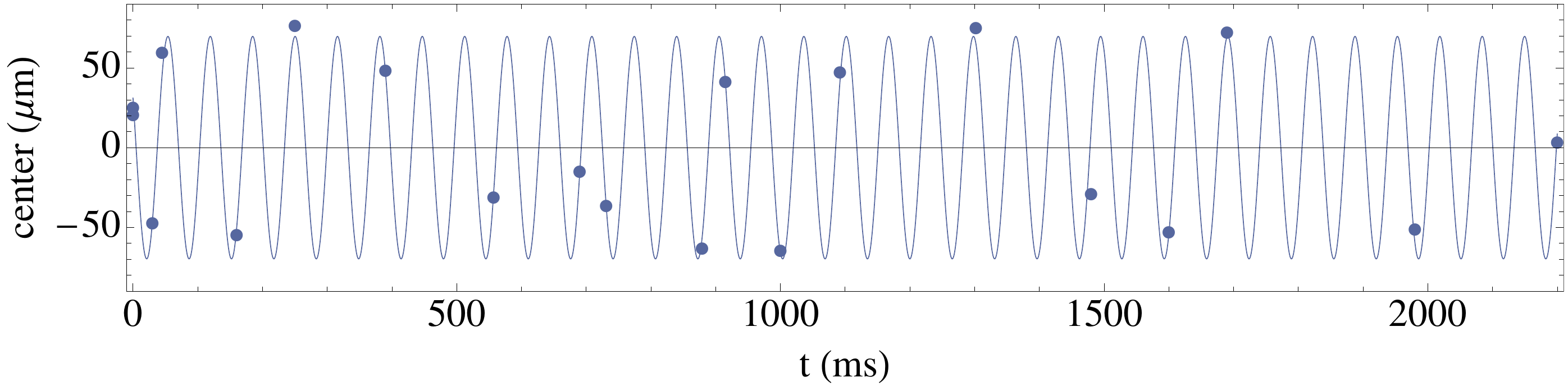} \put(110,15){(a)}\end{overpic}}
\centerline{
\begin{overpic}[width=.6\columnwidth]{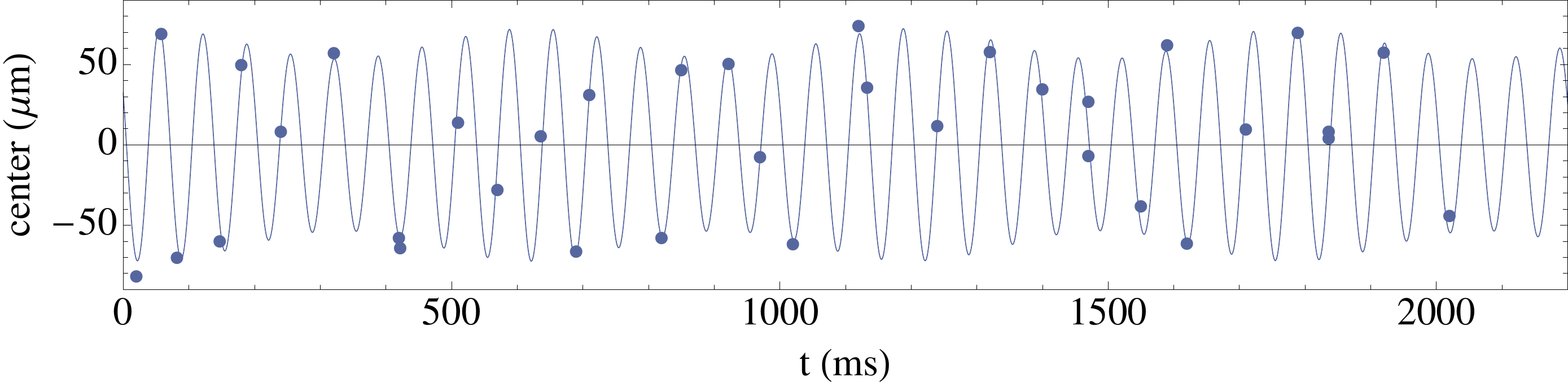} \put(110,15){(b)}\end{overpic}}
\centerline{
\begin{overpic}[width=.6\columnwidth]{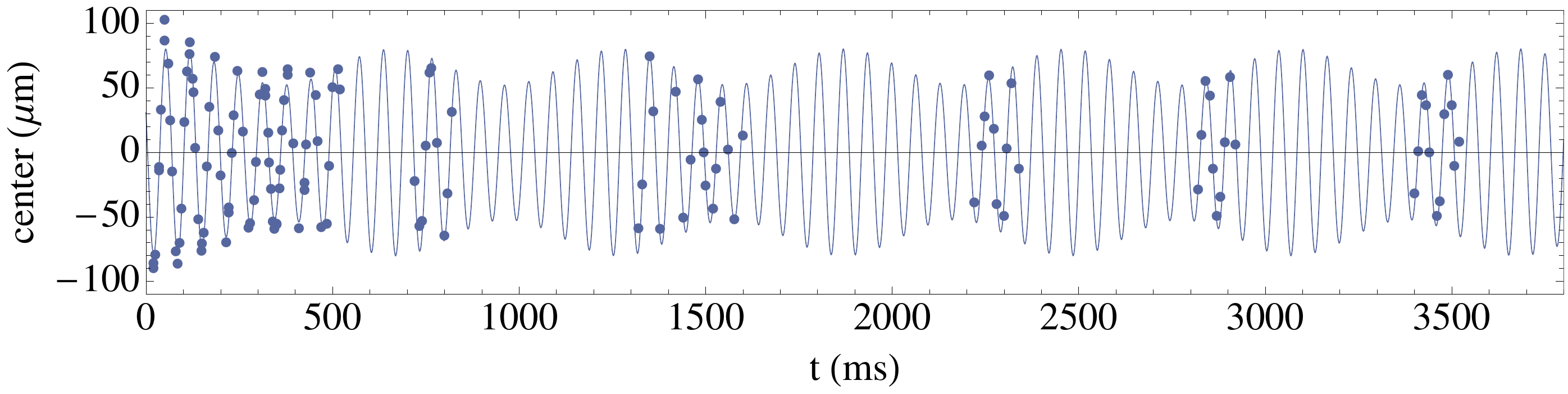} \put(110,15){(c)}\end{overpic}}
\centerline{
\begin{overpic}[width=.6\columnwidth]{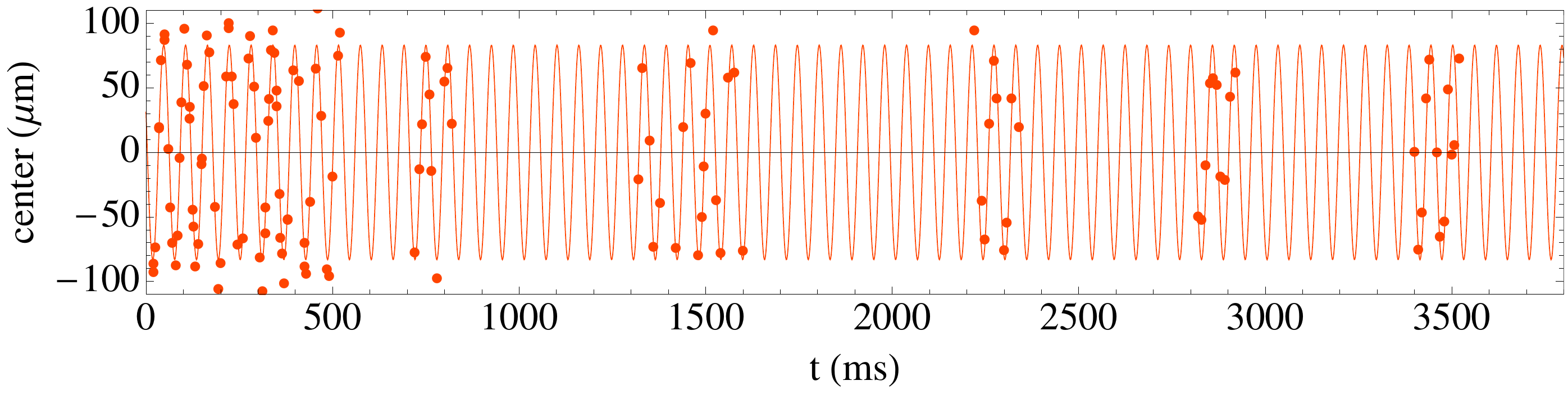} \put(110,15){(d)}\end{overpic}}
\caption{Examples of center-of-mass oscillations.  \7 bosons alone at 832\,G (a), \7 bosons in the presence of \6 fermions at 832G (b).  4 second time span for the  evolution  of \7 bosons (c) mixed with \6 fermions (d) at 835\,G. The coherent energy exchange between \7 and \6 superfluids with no detectable damping is clearly visible.}
\label{Fig6}
\end{figure}

Typical atoms numbers are $N_{\rm b}=4\times10^4$ \7 atoms and $N_{\rm f}=3.5\times10^5$ \6 in a spin-balanced mixture.
The critical temperature for \7 Bose-Einstein condensation is $T_{\rm c,b}=\frac{\hbar\bar \omega_{\rm b}}{k_{\rm B}}\left(N_{\rm b}/\zeta(3)\right)^{1/3}=260$\,nK and the Fermi temperature for \6 $T_{\rm F}=\frac{\hbar\bar \omega_{\rm f}}{k_{\rm B}}\left(3N_{\rm f}\right)^{1/3}=880$\,nK. To our experimental precision, the condensed fraction $\frac{N_0}{N}$ is higher than 0.8, implying $\frac{T_{\rm b}}{T_{\rm c,b}}\lesssim0.5$. With $T_{\rm f}\lesssim T_{\rm b}$, we have $\frac{T_{\rm f}}{T_{\rm F}}\lesssim 0.15=0.8\,T_{\rm c,f}$. This temperature upper bound indicates fermionic superfluidity, in agreement with the direct observation of the superfluid core in the spin-imbalanced gas shown in Fig.~1 in the main text and the extremely low damping observed for small relative oscillations between both isotopes. \\
The large imbalance in isotope population $N_{\rm f}/N_{\rm b}\simeq10$ results from our cooling strategy. At the cost of a small loss in \6 numbers, we can also get samples containing $N_{\rm b}\simeq N_{\rm f,\uparrow}\simeq N_{\rm f,\downarrow}\simeq10^5$.\\
To excite the dipole mode of the two superfluids, we take advantage of the fact that the axial position of the waist of the dipole trap laser beam is slightly off-centered with respect to the minimum of the axial magnetic confinement. In order to displace the center of the atomic clouds, we slowly increase the laser power of the dipole trap by a variable factor (between 1.1 and 2). This results in axial displacement and radial compression of both clouds. The intensity ramp is done in $t_{\rm up}=150$~ms, i.e slow compared to the trap periods. We then return the laser power to its initial value in $t_{\rm down}=20$~ms, fast compared to the axial trap period but slow compared to the radial period, avoiding excitation of radial collective modes. The center of mass positions of both clouds are measured by recording in situ images at variable delays after the axial excitation, up to 4 seconds.  Examples of center-of-mass oscillations over a time span of more than 3.5\,s are shown in Fig \ref{Fig6}.\\

\vspace{.5cm}

{\bf{\large  Critical velocity measurement}}\\

\begin{figure}[htbp]
\centerline{
\includegraphics[width=.6\columnwidth]{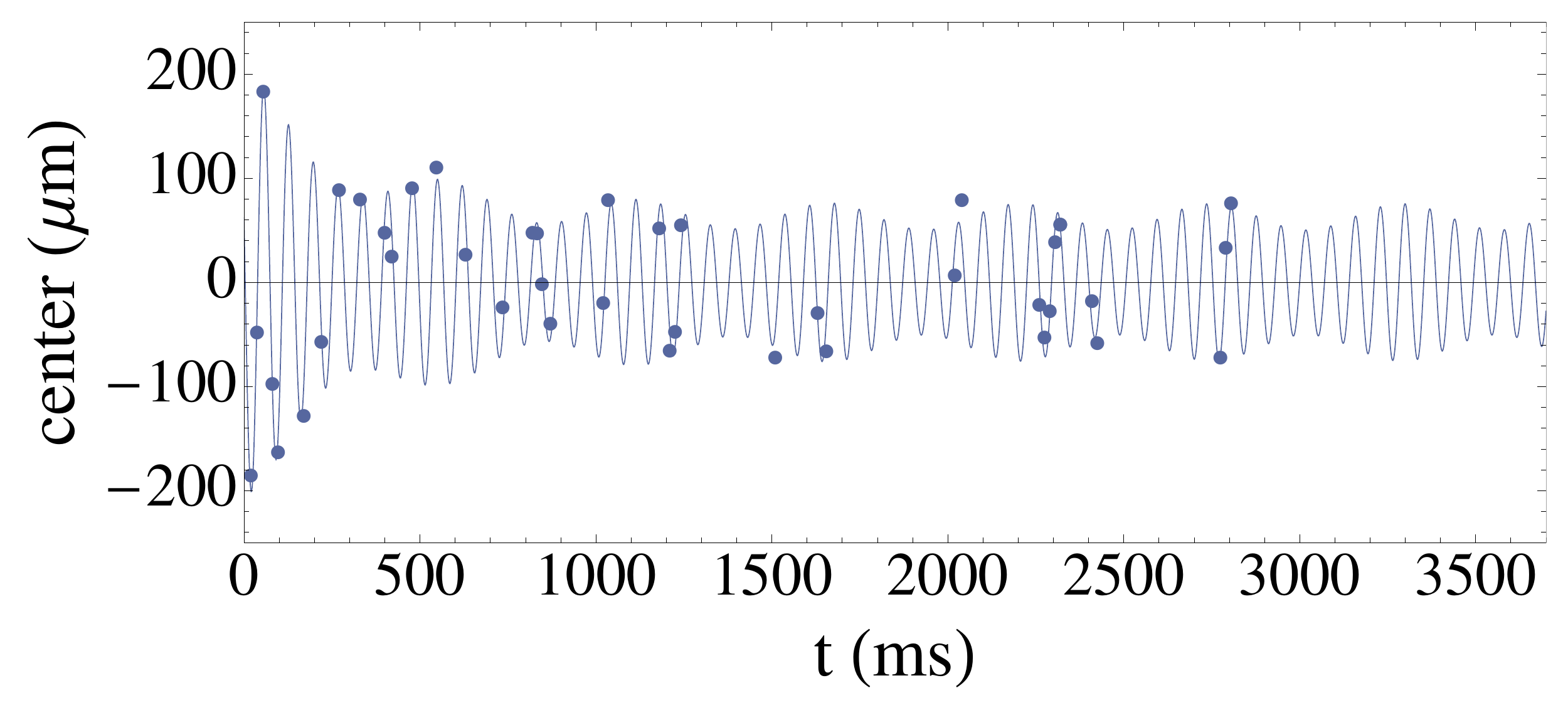}}
\caption{Example of dipole oscillations of the \7 BEC for a large initial amplitude (blue circles). The blue solid line is a fit to equation (7) from main text with a phenomenological damping rate $\gamma= 3.1 {\rm s}^{-1}$.}
\label{damposc}
\end{figure}

When we increase the initial amplitude $d_0$ of the oscillations above $\simeq 100\,\mu$m, we observe first strong damping of the \7 BEC oscillations inside the Fermi cloud followed by long-lived oscillations at a lower amplitude, as shown in Fig. \ref{damposc}.
To verify that this damping is not due to trap anharmonicity for large displacements, we measured oscillations of the BEC in the absence of fermions.  For a displacement of $d=120\,\mu$m, which corresponds to a velocity of $v\simeq0.45v_{\rm F}$ in the presence of the Fermi cloud, we found a characteristic damping rate of $\gamma=0.05\,$s$^{-1}$. For a much larger initial displacement $d=275\,\mu$m ($v\simeq v_{\rm F}$) we observe an influence of trap anharmonicity with an effective damping rate $\gamma=0.26\,$s$^{-1}$. Both of these rates are much smaller than the measured rates in presence of the Fermi cloud for velocities above $0.4\,v_{\rm F}$ as shown in Fig.~2(c) in main text.\\
The observed behavior is compatible with a critical velocity $v_{\rm c}$ for relative motion, resulting in damping for velocities above $v_{\rm c}$ at early times and then undamped oscillations when the velocity is smaller than $v_{\rm c}$. We fit our data with Eq.~(7) from main text and an amplitude $d\,=\,d_0\,{\rm exp}(-\gamma t)+d'$ where $d'$ is the amplitude for the long-lived final oscillations. $\gamma$ is then a damping rate extracted from each data set. Its variation against maximal relative velocity between the two clouds is shown in Fig.~2(c) of main text. To extract a critical velocity we use a simple model:
\bea
\gamma(v)=\Theta(v-v_{\rm c})\,A\,\left((v-v_{\rm c})/v_{\rm F}\right)^\alpha \label{vcModel}
\eea
where $\Theta(x)$ is the Heaviside function, $A$ and $\alpha$ are free parameters.
By fitting Eq.~(\ref{vcModel}), we obtain a critical velocity $v_{\rm c}=0.42^{+0.05}_{-0.11}\,v_{\rm F}$, an exponent $\alpha = 0.95^{+0.8}_{-0.3}$, close to 1, and $A=17(9)\,{\rm s}^{-1}$. This function is plotted in solid blue curve in Fig.~2(c) of main text. $v_{\rm c}$ is very close to the sound velocity of an elongated Fermi gas  $v_{\rm s}'=\frac{\xi^{1/4}}{\sqrt{5}}v_{\rm F}\,=0.35\,v_{\rm F}$ ({\it 21}). For comparison, in a nearly isotropic trap and a moving 1D lattice, the MIT group found a critical velocity $v_{\rm c}=0.32\,v_{\rm F}$ \cite{Miller:2007}.\\


\vspace{.5cm}

{\bf{\large  BEC mean-field and Lee-Huang-Yang limit}}\\

\begin{figure}[h]
\centerline{
\includegraphics[width=.6\columnwidth]{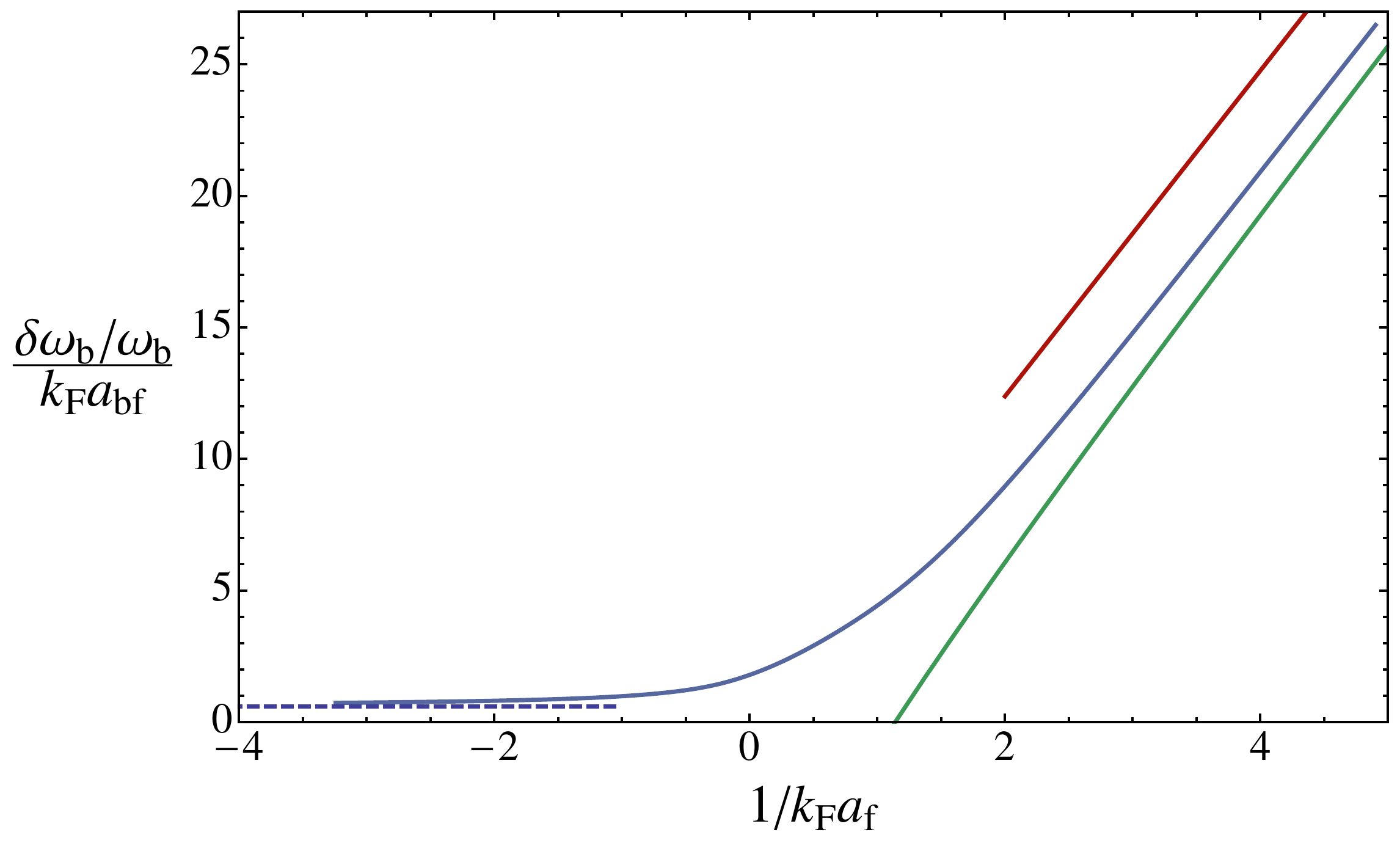}}
\caption{Predicted frequency shift (blue line) over a broad range of $1/k_{\rm F}a_{\rm f}$. The dashed blue line shows the ideal Fermi gas limit. On the BEC side the green line shows the Lee-Huang-Yang prediction (\ref{LHY}) and the red line the mean-field prediction.}
\label{FigLHY}
\end{figure}

Here we evaluate the frequency shift $\delta\omega_{\rm b}/\omega_{\rm b}$ given by 
\be
\frac{\delta\omega_{\rm b}}{\omega_{\rm b}}\simeq\frac12g_{\rm bf}\left(\frac{dn_{\rm f}^{(0)}}{d\mu_{\rm f}}\right)_{r=0},\label{Eq:FshiftSM}
\ee
in the limit where the Fermi superfluid is a molecular BEC of composite Fermi-Fermi dimers. The dimers have a mass $m_{\rm d}=2m_{\rm f}$ and a binding energy $E_{\rm d}=\hbar^2/m_{\rm d}a_{\rm d}^2$, where $a_{\rm d}=0.6\,a_{\rm f}$ is the dimer-dimer scattering length \cite{petrov2004weakly}. The Lee-Huang-Yang EoS for the molecular BEC reads
\be
n_{\rm d}=\frac{\mu_{\rm d}}{g_{\rm d}}\left(1-\frac{32}{3\sqrt{\pi}}\sqrt{\frac{\mu_{\rm d}a_{\rm d}^3}{g_{\rm dd}}}\right)
\ee
where $n_{\rm d}=n_{\rm f}/2$ is the density of dimers, $\mu_{\rm d}=2\mu_{\rm f} +E_{\rm d}$ their chemical potential, and $g_{\rm dd}=4\pi\hbar^2a_{\rm d}/m_{\rm d}$ the coupling constant for the dimer-dimer interaction. Then we have $\frac{d}{d\mu_{\rm f}}=2\frac{d}{d\mu_{\rm d}}$ and thus
\be
\frac{dn_{\rm f}^{(0)}}{d\mu_{\rm f}}=\frac{4}{g_{\rm d}}\left(1-\frac{16}{\sqrt{\pi}}\sqrt{\frac{\mu_{\rm d}a_{\rm d}^3}{g_{\rm dd}}}\right).\label{Eq:FshiftSM2}
\ee
This quantity must be evaluated in the center of the trap ($r=0$) to infer the frequency shift (\ref{Eq:FshiftSM}).
The second term in (\ref{Eq:FshiftSM2}) is of first order in $\sqrt{n_{\rm d}a_{\rm d}}$. We then evaluate its argument in the mean-field approximation which gives the usual expression for the chemical potential of a BEC in a harmonic trap: 
\be
(\mu_{\rm d})_{r=0}=\frac{\hbar\bar\omega_{\rm f}}{2}\left(15N_{\rm d}a_{\rm d}\sqrt{\frac{m_{\rm d}\bar\omega_{\rm f}}{\hbar}}\right)^{2/5}.\label{S7}
\ee
Using (\ref{S7}) and the expression of the Fermi wave-vector:
\be
k_{\rm F}=\sqrt{\frac{m_{\rm d}\bar\omega_{\rm f}}{\hbar}}(6N_{\rm_d})^{1/6},	
\ee
with $N_{\rm f}=2N_{\rm d}$, we can recast our expression for the frequency shift (\ref{Eq:FshiftSM2}) in the universal units used in the main text (Eq.~(10)): 
\bea
\left(\frac{\mu_{\rm d}a_{\rm d}^3}{g_{\rm d}}\right)_{r=0}=\frac{1}{8\pi}\left(\frac 52\right)^{2/5}\left(a_{\rm d}k_{\rm F}\right)^{12/5}\\
\left(\frac{dn_{\rm f}^{(0)}}{d\mu_{\rm f}}\right)_{r=0}\simeq\frac{2m_{\rm f}}{0.6\pi\hbar^2a_{\rm f}}\left(1-1.172\left(k_{\rm F}a_{\rm f}\right)^{6/5}\right)\\
\frac{\delta\omega_{\rm b}}{\omega_{\rm b}}\frac{1}{k_{\rm F}a_{\rm bf}}\simeq6.190\frac{1}{k_{\rm F}a_{\rm f}}\left(1-1.172\left(k_{\rm F}a_{\rm f}\right)^{6/5}\right)\label{LHY}
\eea 
This limit is shown in green in Fig. \ref{FigLHY}. The mean-field approximation (red curve in Fig. \ref{FigLHY}) corresponds to the first term in Eq.~(\ref{LHY}).\\

\newpage

{\bf{\large  Derivation of the coupled oscillator model using the sum-rule approach}}\\

We describe the dynamics of the system by a Hamiltonian

\be
\widehat H=\sum_{i,\alpha}\left[\frac{\widehat p_{\alpha,i}^2}{2m_\alpha}\right]+U(\bm r_{\alpha,i}),
\ee
where $\alpha={\rm b,f}$ labels the isotopes, and $U$ describes the total (trap+interaction) potential energy of the cloud.

Consider the operators $\widehat F_\alpha=\sum_{i=1}^{N_\alpha} \widehat z_{\alpha,i}$, where  $z_{\alpha,i}$ is the position along $z$ of the $i$-th atom of species $\alpha={\rm b,f}$ and take $\widehat F(a_{\rm f},a_{\rm b})=\sum_\alpha a_\alpha \widehat F_\alpha$ an excitation operator depending on two mixing coefficients $(a_\alpha)$. We introduce the moments $S_p$ defined by
$$
S_p=\sum_n (E_n-E_0)^p\left|\langle n|\widehat F|0\rangle\right|^2,
$$
where $|n\rangle$ and $E_n$ are the eigenvectors and the eigenvalues of the Hamiltonian $\widehat H$ (by definition $|0\rangle$ is the ground state and $E_0$ is its energy). Using the Closure Relation and first order perturbation theory, $S_1$ and $S_{-1}$ can be calculated exactly and we have
\begin{eqnarray}
S_1&=&-\sum_{\alpha} \frac{\hbar^2}{m_\alpha} N_\alpha a_\alpha^2\label{Eq:sumrule1}\\
S_{-1}&=&-\frac{1}{k}\sum_{\alpha,\beta}a_\alpha a_\beta N_\alpha\frac{\partial \langle z_\alpha\rangle}{\partial b_\beta}\label{Eq:sumrule2}
\end{eqnarray}
where $k$ is the restoring force of the axial magnetic trap and $\langle z_\alpha\rangle$ is the center of mass position of atoms $\alpha$ in the presence of a perturbing potential $-k\sum_\beta b_\beta \widehat F_\alpha$ corresponding to a shift of the trapping potential of species $\beta$ by a distance $b_\beta$.  $\langle z_\alpha\rangle$ satisfies two useful conditions. First, using Hellmann-Feynman's theorem, the matrix $N_\alpha\partial_{b_\beta}\langle z_\alpha\rangle=\partial^2_{b_\alpha b_\beta} \widehat H$ is symmetric. Secondly, if we shift the two traps by the same quantify $b_\beta = b$, the center of mass of the two clouds move by $\langle z_\alpha\rangle=b$. Differentiating this constraint with respect to $b$ yields the condition $\sum_\beta \partial_{b_\beta}\langle z_\alpha\rangle=1$.

Experimentally, we observe that only two modes are excited by the displacement of the trap center. We label $|n=1\rangle$ and $|n=2\rangle$ the corresponding modes and we take $\hbar\omega_n=E_n-E_0$, with, by convention, $\omega_1\le \omega_2$. We thus have for any set of mixing parameters $(a_{\rm f},a_{\rm b})$,
\be
\hbar^2\omega_1^2\le \frac{S_1}{S_{-1}}\le \hbar^2\omega_2^2.
\ee
To find the values of the two frequencies $\omega_1$ and $\omega_2$, one thus simply has to find the extrema of $S_1/S_{-1}$ with respect to $a_{\rm f}$ and $a_{\rm b}$. Using the sum rules (\ref{Eq:sumrule1}) and (\ref{Eq:sumrule2}), we see that
\be
\frac{S_1}{S_{-1}}=\hbar^2 k\frac{\sum_{\alpha}N_\alpha/m_\alpha a_\alpha^2}{\sum_{\alpha,\beta}N_\alpha a_\alpha a_\beta \frac{\partial \langle z_\alpha\rangle}{\partial b_\beta}}.
\ee
This expression can be formally simplified by taking $a'_\alpha=a_\alpha \sqrt{N_\alpha/m_\alpha}$ and $\psi=(a'_{\rm f},a'_{\rm b})$. We then have
\be
\frac{S_1}{S_{-1}}=\hbar^2k\frac{\langle\psi|\psi\rangle}{\langle\psi|{\cal M}\psi\rangle},
\ee
where the scalar product is defined by $\langle\psi|\psi'\rangle=\sum_\alpha \psi_\alpha\psi'_\alpha$ and the effective-mass operator is given by
\be
{\cal M}_{\alpha\beta}=\sqrt{m_\alpha m_\beta}\sqrt{\frac{N_\alpha}{N_\beta}}\frac{\partial\langle z_\alpha\rangle}{\partial b_\beta}.
\ee
With these notations, the frequencies $\omega_{i=1,2}$ are given by $\omega_i=\sqrt{k/\tilde m_i}$, where $\tilde m_i$ is an eigenvalue of ${\cal M}$.

In the weak-coupling limit, the cross-terms $\partial_{b_\beta} \langle z_\alpha\rangle$ ($\alpha\not = \beta$) are small and using their symmetry properties, we can write ${\cal M}$ as ${\cal M}_0+{\cal M}_1$ with
\footnotesize
\begin{eqnarray}
&{\cal M}_0=\left(\begin{array}{cc}
m_{\rm f}&0\\
0&m_{\rm b}
\end{array}\right)
\\
&{\cal M}_1=\left(\begin{array}{cc}
-m_{\rm f}\frac{\partial \langle z_{\rm f}\rangle}{\partial b_{\rm b}}&\sqrt{m_{\rm f}m_{\rm b}}\sqrt{\frac{N_{\rm b}}{N_{\rm f}}}\frac{\partial\langle z_{\rm b}\rangle}{\partial b_{\rm f}}\\
\sqrt{m_{\rm f}m_{\rm b}}\sqrt{\frac{N_{\rm b}}{N_{\rm f}}}\frac{\partial\langle z_{\rm b}\rangle}{\partial b_{\rm f}}&-m_{\rm b}\frac{\partial \langle z_{\rm b}\rangle}{\partial b_{\rm f}}
\end{array}\right)
\end{eqnarray}
\normalsize
Since the matrix $\cal M$ is symmetric we can use the usual perturbation theory to calculate its eigenvalues and eigenvectors. We have to first order

\begin{eqnarray}
\tilde m_1&=&m_{\rm f}\left(1-\frac{\partial\langle z_{\rm f}\rangle}{\partial b_{\rm b}}\right)\\
\tilde m_2&=&m_{\rm b}\left(1-\frac{\partial\langle z_{\rm b}\rangle}{\partial b_{\rm f}}\right)
\end{eqnarray}

Using the symmetry of $N_\alpha\partial_{b_\beta}\langle z_\alpha\rangle$, we see that in the experimentally relevant limit $N_{\rm f}\gg N_{\rm b}$, we have $\partial_{b_{\rm f}}\langle z_{\rm b}\rangle\gg \partial_{b_{\rm b}}\langle z_{\rm f}\rangle$. Thus the frequency of $^6$Li is essentially not affected by the coupling between the two species. To leading order, we can identify $\omega_1$ ($\omega_2$) with $\tilde\omega_{\rm b}$ ($\tilde\omega_{\rm f}$) and we have

\begin{eqnarray}
\tilde\omega_{\rm f}&\simeq&\omega_{\rm f}\\
\tilde\omega_{\rm b}&\simeq&\omega_{\rm b}\left(1+\frac{1}{2}\frac{\partial \langle z_{\rm b}\rangle}{\partial b_{\rm f}}\right)\label{Eq:FreqShift}
\end{eqnarray}

To calculate the frequency $\tilde\omega_{\rm b}$ we need to know the crossed-susceptibility $\partial_{b_{\rm f}}\langle z_{\rm b}\rangle$. Since this is in equilibrium quantity, we can calculate it using the local-density approximation. We then obtain
\be
\frac{\partial \langle z_{\rm b}\rangle}{\partial b_{\rm f}}=\frac{kg_{\rm bf}}{N_{\rm b}}\int d^3\bm r  z^2\left(\frac{\partial n_{\rm f}}{\partial\mu_{\rm f}}\right)\left(\frac{\partial n_{\rm b}}{\partial\mu_{\rm b}}\right)
\ee
In the limit $N_{\rm b}\ll N_{\rm f}$, the bosonic cloud is much smaller than the fermionic cloud. We can therefore approximate this expression by
\be
\frac{\partial \langle z_{\rm b}\rangle}{\partial b_{\rm f}}\simeq\frac{kg_{\rm bf}}{N_{\rm b}}\left(\frac{\partial n_{\rm f}}{\partial\mu_{\rm f}}\right)_0\int d^3\bm r  z^2\left(\frac{\partial n_{\rm b}}{\partial\mu_{\rm b}}\right)
\ee
where the index zero indicates that the derivative is calculated at the center of the trap. The integral can be calculated exactly and we finally obtain
\be
\frac{\partial \langle z_{\rm b}\rangle}{\partial b_{\rm f}}=g_{\rm bf}\left(\frac{\partial n_{\rm f}}{\partial \mu_{\rm f}}\right)_0,
\ee
where we recover Eq.~(2) from main text.

To get the dynamics of the system after the excitation, we need to calculate the eigenvectors of the matrix $\cal M$. Note $\psi'_i=(a'_{i,{\rm f}},a'_{i,{\rm b}})$ the eigenvector associated to the eigenvalue $\omega_i$. Using once more first order perturbation theory, we have
\begin{eqnarray}
\psi'_1=\left(
\begin{array}{c}1\\
\frac{\sqrt{m_{\rm f}m_{\rm b}}}{m_{\rm f}-m_{\rm b}}\sqrt{\frac{N_{\rm b}}{N_{\rm f}}}\frac{\partial\langle z_{\rm b}\rangle}{\partial b_{\rm f}}
\end{array}\right)\\
\psi'_2=\left(\begin{array}{c}
\frac{\sqrt{m_{\rm f}m_{\rm b}}}{m_{\rm b}-m_{\rm f}}\sqrt{\frac{N_{\rm b}}{N_{\rm f}}}\frac{\partial\langle z_{\rm b}\rangle}{\partial b_{\rm f}}\\
1
\end{array}\right),
\end{eqnarray}
 from which we deduce the vectors $\psi_{i=1,2}=(a_{i,{\rm f}},a_{i,{\rm b}})$ giving the excitation operator $\widehat F(a_{i,{\rm f}},a_{i,{\rm b}})$. More precisely
\begin{eqnarray}
\psi_1&=&\sqrt{\frac{m_{\rm f}}{N_{\rm f}}}\left(
\begin{array}{c}1\\
\frac{m_{\rm b}}{m_{\rm f}-m_{\rm b}}\frac{\partial\langle z_{\rm b}\rangle}{\partial b_{\rm f}}
\end{array}\right)
\\
\psi_2&=&\sqrt{\frac{m_{\rm b}}{N_{\rm b}}}\left(\begin{array}{c}
\frac{m_{\rm f}}{m_{\rm b}-m_{\rm f}}\frac{N_{\rm b}}{N_{\rm f}}\frac{\partial\langle z_{\rm b}\rangle}{\partial b_{\rm f}}\\
1
\end{array}\right).
\end{eqnarray}

Note $d$ the initial displacement of the two species and  expand the initial condition $Z=(z_{\rm f}(0),z_{\rm b}(0))=(d,d)$ over the basis $\{\psi_1,\psi_2\}$ as $Z=\sum_i c_i \psi_i$.   Since by construction the operator $\widehat F(a_{i,{\rm f}},a_{i,{\rm b}})$ excites solely the mode $\omega_i$ we must have at time $t$ $Z(t)=\sum_i c_i \cos(\omega_i t)\psi_i$ (we assume that the initial velocities are zero). After a straightforward calculation, we get
\footnotesize
\begin{eqnarray}
z_{\rm f}(t)&=&d\left[\frac{(1-\varepsilon\rho\eta)\cos(\omega_1 t)+\eta\rho\varepsilon(1+\varepsilon)\cos(\omega_2 t)}{1+\varepsilon^2\rho\eta}\right]\\
z_{\rm b}(t)&=&d\left[\frac{-\varepsilon(1-\varepsilon\rho\eta)\cos(\omega_1 t)+(1+\varepsilon)\cos(\omega_2 t)}{1+\varepsilon^2\rho\eta}\right]
\end{eqnarray}
\normalsize
with $\rho=N_{\rm b}/N_{\rm f}$, $\varepsilon=m_{\rm b}/(m_{\rm b}-m_{\rm f})\partial_{b_{\rm f}}\langle z_{\rm b}\rangle$ and $\eta=m_{\rm f}/m_{\rm b}$. In experimentally relevant situations, we have $\varepsilon\ll 1$, $\rho\ll 1$ and $\eta\simeq 1$. We can thus approximate the previous equations by
 \begin{eqnarray}
z_{\rm f}(t)&\simeq&d\left[(1-\varepsilon\rho)\cos(\tilde\omega_{\rm f} t)+\rho\varepsilon\cos(\tilde\omega_{\rm b} t)\right]\\
z_{\rm b}(t)&\simeq&d\left[-\varepsilon\cos(\tilde\omega_{\rm f} t)+(1+\varepsilon)\cos(\tilde\omega_{\rm b} t)\right],
\end{eqnarray}
 and where according to Eq. (\ref{Eq:FreqShift}), we can take
 \be
 \varepsilon =\frac{2m_{\rm b}}{m_{\rm b}-m_{\rm f}}\left(\frac{\tilde\omega_{\rm b}-\omega_{\rm b}}{\omega_{\rm b}}\right).
 \ee\\
 \vspace{-.2cm}
\bibliographystyle{Science}

\end{document}